\documentclass[11pt,twoside]{article}
\usepackage{asp2006}
\usepackage{lscape}
\usepackage{graphics}
\usepackage{graphicx}
\newcommand{\ha}{H$\alpha$~}

\newcommand{\kpc}{{\rm kpc}}
\newcommand{\kms}{\,km\,s$^{-1}$}

\newcommand{\myr}{\,$M_{\sun}\,{\rm yr}^{-1}$}

\newcommand{\ecsa}{$\rm\,erg\,cm^{-2}\,s^{-1}\,\AA^{-1}$}
\begin{document}
\title{A non-spherical mass outflow from RS~Oph 
       during its 2006 outburst}
\author{A. Skopal, T. Pribulla}
\affil{Astronomical Institute, Slovak Academy of Sciences,
       059\,60 Tatransk\'a Lomnica, Slovakia}
\author{Ch. Buil}
\affil{Castanet Tolosan Observatory, 6 Place Clemence Isaure,
       31320 Castanet Tolosan, France}
\author{A. Vittone, L. Errico}
\affil{INAF Osservatorio Astronomico di Capodimonte,
       via Moiariello 16, I-80 131 Napoli, Italy}
\begin{abstract} %%% Abstract to run on from here.
We present results of our modeling the \ha line profile 
along the 2006 RS\,Oph outburst. At day 1.38 the very broad 
component of the \ha profile was possible to fit by a bipolar 
wind model. The model corresponds to a very fast acceleration 
of the wind particles and the line luminosity of 
$\sim 2\,900 (d/1.6\kpc)^2\,L_{\odot}$ to the mass-loss 
rate of $\sim (1-2)\times 10^{-4}$\myr. 
During days $10-30$ the broad component shrank 
to $FWZI \sim 1\,800$\kms. It could be associated with 
expanding ring and its satellite components at 
$\sim\pm 2\,430$\kms\ with bipolar jets. 
Later observations made at day 57 and 209 indicated 
a decrease in both the mass-loss rate 
($\sim 1\times 10^{-5} - 1\times 10^{-6}$\myr) and the wind 
acceleration. During the quiet phase, emission bumps observed 
sporadically in the line wings could reflect clumpy ejections 
by the central star. 
\end{abstract}
%
%%% MAIN BODY OF TEXT GOES HERE. CONSULT "INSTRUCTIONS FOR AUTHORS USING
%%% LATEX2E MARKUP", SECTIONS 2.3-2.6 FOR HELP WITH EQUATIONS, FIGURES,
%%% AND TABLES.
%
\section{Introduction}

RS\,Oph is a symbiotic recurrent nova, in which a high-mass 
white dwarf (WD) accretes material from a cool K7\,III giant 
on a 456-day orbit \citep[e.g.][]{bode87,ms99,f+00}. 
Historically, 6 eruptions have been recorded unambiguously. 
The first one in 1898 and the last one on 2006 February 12.83 
\citep[][]{evans88, nar+06}. 
The recurrence period of approximately 20 years and a bright 
peak magnitude, $V = 4-5$, made RS\,Oph a good target for 
multifrequency observational campaigns from the beginning 
of its recent, 1985 and 2006, outbursts 
\citep[e.g.][and references therein]{evans07}. 
For example, they revealed a non-spherical shaping of the 
nova ejecta: (i) on the radio map \cite[e.g.][]{taylor+89}, 
(ii) by interferometric technique in the near infrared 
\citep[e.g.][]{lane+07,chesneau+07} and also (iii) by 
the HST imaging in the optical \citep{bode+07}. 

In this contribution we present evidences of the non-spherical
mass outflow on the basis of the optical spectroscopy
of RS\,Oph carried out from the first days of its 2006 outburst 
to the post-outburst quiet phase. 
%
%==============================================|
%---- Fig. 1: H-alpha 1.38 + beta-law wind ----|
%==============================================|
%
\begin{figure}[!t]
\centering
\begin{center}
\resizebox{\hsize}{!}{\includegraphics[angle=-90]{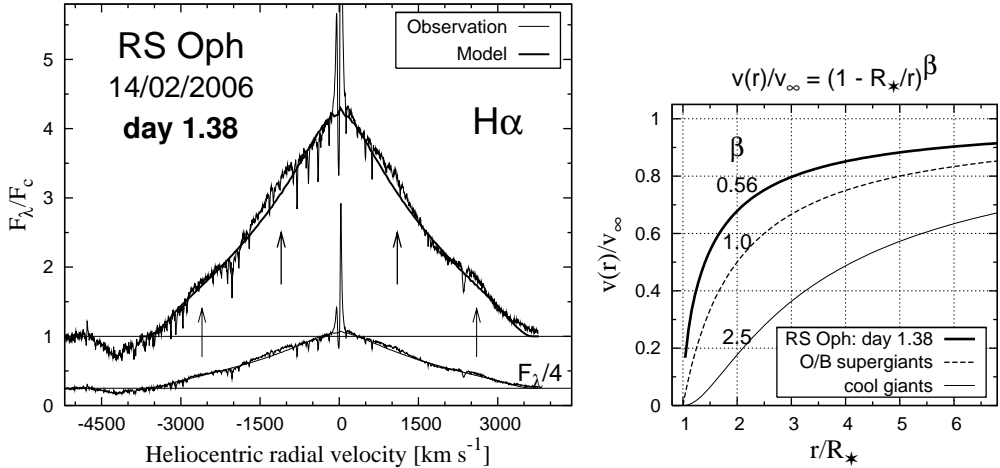}}
\caption[]{Left: \ha line observed at day 1.38 at the Castanet 
Toloson Observatory with resolution of 0.115\AA/pixel. The broad 
component, triangular in profile, can be matched by a bipolar 
wind model with a very high acceleration of the particles. 
Arrows denote extra emissions located symmetrically around 
$\pm$1\,000 and $\pm$2\,500\kms. The level of the continuum 
corresponds to $1.5\times 10^{-10}$\ecsa. Right: A comparison 
of the $\beta$ factor in the wind-law for different types of 
the stars. 
          }
\end{center}
\end{figure}

\section{Bipolar wind at day 1.38}

The \ha profile from day 1.38 consisted of a very broad 
component, triangular in profile, with $FWZI\sim 7\,600$\kms, 
a narrow emission component ($FWZI\sim 240$\kms) cut with 
a sharp absorption at $\sim -13$\kms\ and a blueward-shifted 
absorption component at $\sim -4\,250$\kms\ (Fig.~1). 
% suggesting a P-Cyg type of the profile (Fig.~1). 

We suggest that the broad triangular profile is due to 
kinematics of the photoionized and optically thin stellar 
wind from the WD. The narrow central absorption/emission 
components then could represent result of the radiative 
transfer in \ha through the optically thick fraction of 
the wind at the vicinity of the WD surface. 
The violet-shifted absorption could be associated with a dense 
material of the neutral wind from the giant swept off by 
the fast wind from the WD and accumulated into a shell at 
the front of the ejecta. 
%
% OR: The fast wind sweeps up the slower moving neutral wind from 
%the giant into a shell at the front of the ejecta,

We tested the origin of the broad triangular component 
by the bipolar wind model as proposed by \cite{sk06}. 
In this model a fraction of the wind from the central star 
is blocked by the disk at/around the WD's equator, which 
thus creates bipolar geometry of the stellar wind. 
According to the model SED from the quiet phase \citep{sk+08} 
we adopted the disk radius $R_{\rm D} = 10\,R_{\odot}$ 
and the disk thickness at its edge, $H = 3.3\,R_{\odot}$. 
%
%assuming a flared disk with $H/R_{\rm D} = 0.3$. 
%
The resulting synthetic profile (Fig.~1, left) corresponds 
to the terminal velocity $v_{\infty} = (3\,800 \pm 100)$\kms, 
the acceleration parameter $\beta = 0.56\pm 0.02$ and 
the mass loss rate through the wind, 
$\dot M_{\rm W} = (1-2)\times\,10^{-4}$\myr. 
The small value of $\beta$ reflects a very high acceleration 
of the wind particles (Fig.~1, right).
%
%Right panel of Fig.~1 compares 
%this quantity to other types of the stars. 
%
%Not possible to fit the profile with a spherically symm. wind 
%- bipolar wind model: parameters
%
%==============================================|
%---- Fig. 2: H-alpha day 12.4 + 15.3: Jets----|
%==============================================|
%
\begin{figure}[!t]
\centering
\begin{center}
\resizebox{\hsize}{!}{\includegraphics[angle=-90]{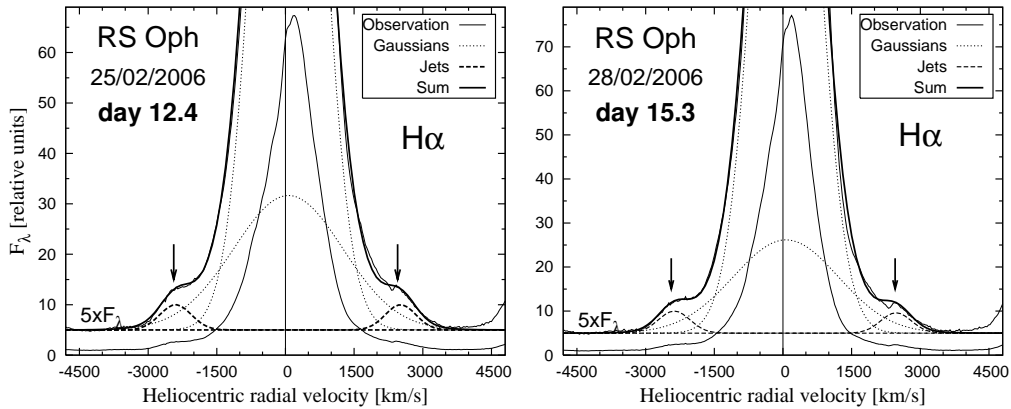}}
\caption[]{
\ha line observed at day 12.4 and 15.3 at the Castanet Toloson 
Observatory (resolution of 0.74\AA/pixel). High-velocity satellite
components (jets) were located at $\sim\pm 2\,430$\kms\ (arrows). 
They were isolated from the total \ha profile by fitting 
the central emission with two Gaussian curves. 
%(dotted curves).
%The level of the continuum corresponds 
%to 2.2 and 1.8$\times 10^{-11}$\ecsa\ at day 12.4 and 15.3, 
%respectively. 
          }
\end{center}
\end{figure}

\section{Bipolar jet-like collimated outflow during 
         $\sim 10-30$ days}

Here we present examples of \ha line profiles we measured 
at day 12.4 and 15.3 (Fig.~2). Similar profile could still 
be recognized at day 28.3 (13/03/2006). The whole profile 
was shifted 
to the red part of the spectrum by $\sim$150\kms. The main 
emission core was triangular in profile with the peak 70--80 
times the continuum and the base of $FWZI = 1\,750 \pm 250$\kms. 
In addition, the profile showed emission shoulders located 
bipolarly at $\sim\pm 2\,430$\kms, expanding to 
$\sim \pm 3\,000$\kms (Fig.~2). 
Radial velocities of both the triangular emission core and 
the faint satellite components in the \ha profile were very 
similar to those indicated by the AMBER/VLTI interferometric 
observations at day 5.5 \citep{chesneau+07}: a slow expanding 
ring-like structure with $|v_{\rm rad}| < 1\,800$\kms and 
a fast structure ($|v_{\rm rad}| \sim 2\,500-3\,000$\kms) 
in the direction that coincides with the jet-like feature 
seen in the radio. 
Accordingly, the triangular \ha emission core could be 
associated with an expanding ring at the equatorial plane 
of the WD and the satellite components with bipolar jets ejected 
at/around the accretor. The latter, however, requires 
the presence of the inner disk, through which sufficient 
amount of mass is accreted to balance the outflow via 
the jets \citep[note that 
$\dot M_{\rm acc}\ga\dot M_{\rm jet}$, e.g.][]{livio+03}. 
%
%On the other hand, if we attribute the satellite components 
%to bipolar jets, then inner disk has to be present to give 
%rise the mass outflow via the jets 
%(note that $\dot M_{\rm acc} \ga \dot M_{\rm jet}$ 
%\cite[e.g.][]{livio+97}). 
%
%\section{Attenuation of the stellar wind during 57--111 days}
%\section{Clumpy ejections during the post-outburst minimum}
%
%======================================|
%---- Fig. 3: H-alpha day 57 + 209 ----|
%======================================|
%
\begin{figure}[!t]
\centering
\begin{center}
\resizebox{\hsize}{!}{\includegraphics[angle=-90]{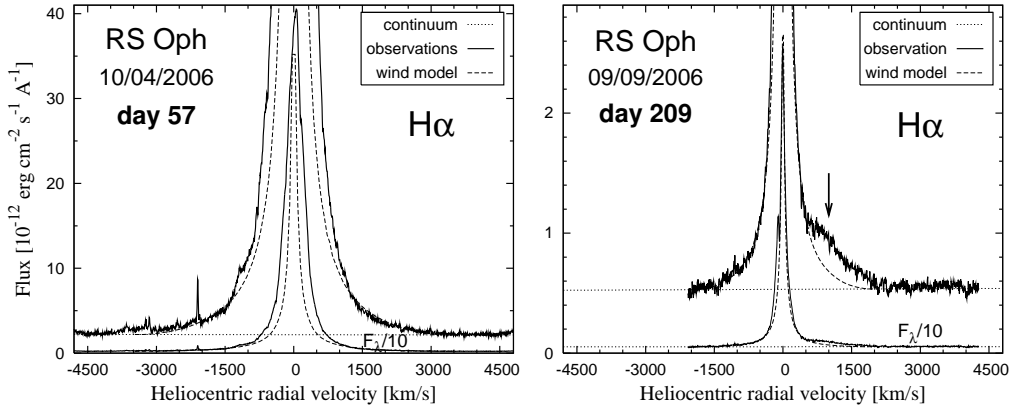}}
\caption[]{\ha line observed at day 57 at the David Dunlop 
Observatory (R=12000) and 209 at the Asiago Astrophysical 
Observatory (R=25000). 
          }
\end{center}
\end{figure}
\section{Attenuation of the stellar wind and clumpy ejections} 

Figure~3 shows \ha line profiles carried out at day 57 
(10/04/2006) and 209 (09/09/2006) compared with synthetic 
profiles of the bipolar wind we introduced in Sect.~2. 
In the earlier case (left panel) the wind model fits only 
the very extended wings ($|\Delta RV| \ga 1\,300$\kms) of 
the observed profile. The model corresponds to 
$v_{\infty}\sim 3\,500$\kms, $\beta \sim 1.7$ and the mass 
loss rate, $\dot M_{\rm W} \sim 1.5\times 10^{-5}$\myr\ 
for $d = 1.6$\,kpc. However, a significant fraction of 
the line core does not obey the $\beta$-wind velocity law. 
The $FWHM \sim 420$\kms\ is too large to be fitted by the 
wind. This reflects the presence of a bipolarly located 
emission within $|\Delta RV| \la 1\,000$\kms\ having probable 
cause in the expanding and diluting ring-like structure 
as discussed in Sect.~3. 

At day 209 the \ha flux decreased by a factor of $\sim$20 and 
the profile became to be narrow ($FWHM \sim 100$\kms). 
The model fits well the profile for $|\Delta RV| \ga 200$\kms, 
but isolates an emission bump in the red wing. We interpret this 
feature as a result of a sporadic clumpy ejection from 
the accretor due to abrupt accretion from the re-creating large 
disk during the quiet phase 
\citep[c.f. the SED in Fig.~1 of][]{sk+08}. 

\acknowledgements
 This work was supported by the Slovak Academy of Sciences
 grant No. 2/7010/7. AS acknowledges the LOC for their support. 
\end{document}